\documentclass[a4paper,singlecoulomb,12pt]{article}
\usepackage{epsfig,amsmath,amssymb}
\usepackage[colorlinks=true,linktocpage=true,linkcolor=blue,citecolor=blue]{hyperref}
\usepackage{cite}
\usepackage{graphicx}
\usepackage{dcolumn}
\usepackage{bm}
\usepackage{subfig}
\usepackage{float}
\usepackage{tabularx}
\usepackage{hhline}
\usepackage{color}
\usepackage{latexsym}
\newcommand{\old}[1]{}

\newcommand{\be}{\begin{equation}}
\newcommand{\ee}{\end{equation}}
\newcommand{\ba}{\begin{eqnarray}}
\newcommand{\ea}{\end{eqnarray}}
\newcommand{\bi}{\begin{itemize}}
\newcommand{\ei}{\end{itemize}}

\newcommand{\nd}{\noindent}

\usepackage{caption}

\begin{document}
\begin{flushright}
{\normalsize
}
\end{flushright}
\vskip 0.1in
\begin{center}
{\large {\bf Effects of strong magnetic field on the formation of wakes in 
thermal QCD}}
\end{center}
\vskip 0.1in
\begin{center}
Mujeeb Hasan\footnote{mhasan@ph.iitr.ac.in} and Binoy Krishna Patra
\footnote{binoy@ph.iitr.ac.in}
\vskip 0.02in
{\small {\it Department of Physics, Indian Institute of
Technology Roorkee, Roorkee 247 667, India}}\\
\end{center}
\vskip 0.01in

\begin{abstract}
We have investigated how the wakes in the induced charge density and in 
the potential due to the passage of highly
energetic partons through a thermal QCD medium 
get affected by the presence of strong magnetic field ($B$). 
For that purpose, we wish to analyze first the dielectric 
responses of the medium both in presence and absence of strong 
magnetic field. Therefore, we have revisited the 
general form for the gluon self-energy 
tensor at finite temperature and finite magnetic field and then 
calculate the relevant structure functions at finite temperature
and strong magnetic field limit (SMF: $|q_f B| \gg T^2$ as well as
$|q_fB| \gg m_f^2$, $q_f (m_f)$ is the electric charge (mass)
of $f$-th flavour). We found that for slow moving partons, the 
real part of dielectric function is not affected by the magnetic 
field whereas for fast moving
partons, for small $|\textbf{k}|$, it 
becomes very large and approaches towards 
its counterpart at $B=0$, for large $|\textbf{k}|$.  On the other 
hand the imaginary part is decreased for both
slow and fast moving partons, due to the fact that the
imaginary contribution due to quark-loop vanishes. 
With these ingredients, we found that the oscillation in 
the (scaled) induced charge density, due to the very fast partons
becomes less pronounced in the presence of strong magnetic field
whereas for smaller parton velocity, no significant change is 
observed. For the (scaled) wake potential along the motion 
of fast moving partons (which is of Lennard-Jones (LJ) type), 
the depth of negative minimum in the backward region gets
reduced drastically, resulting in the reduction of the amplitude of 
oscillation. On the other hand in the forward region, it remains
as the screened Coulomb one, except the screening now becomes
much stronger for higher parton velocity. 
Similarly for the wake potential transverse to the motion
of partons in both forward and backward regions, the depth of 
LJ potential for fast moving partons gets
decreased severely, but still retains the forward-backward symmetry. 
However, for lower parton velocity,
the magnetic field does not affect it significantly. 
\end{abstract}

\vspace{1mm}

\section{Introduction}
Ultra-relativistic heavy-ion collisions (URHICs) at RHIC, e-RHIC, 
LHC, upcoming FAIR etc. provides an enticing opportunity to identify  
the transition from the hadronic matter to the quark matter,
known as quark-gluon plasma (QGP). The hard scatterings at the early stages 
of URHICs produce very high energetic partons, known as jets. The jets
behave as an external hard probe, and while ploughing through the 
medium, it may loose energy and momentum either through the radiation~\cite{Baier:PRD58_1998,
Zakharov:JETP_1997,Baier:JHEP_2001,Kovner:0304151,Gyulassy:NP420_1994,Gyulassy:NPB594_2001} 
or the 
collision~\cite{Mustafa:APHA22_2005,Munshi:PRC_2005} or both~
\cite{Wicks:NPA783_2007,Guang:PRL100_2008,Zakharov:JETP88_2008} 
depending on the range of momentum, 
altogether known as jet quenching~\cite{Gyulassy:PLB243_1990,Wang:PRL89_2002,
Mustafa:APHA22_2005} and have been manifested 
largely by the suppression of high $p_T$ hadronic yields~\cite{Bjorken:Report82_1982}. 
One of the specific consequences of the jet quenching is the dihadron 
azimuthal correlation at RHIC, which is manifested 
in the hadron pair-distribution in the intermediate $p_T$ range 
through a double peak 
structure in the away side~\cite{Adams:PRL95_2005,Adler:PRL97_2006}. 
Although, how the jet-medium interaction 
affects the distribution is not a settled issue, nevertheless
the coupling of jets to a strongly interacting medium either at 
the particle level or at the collective level may explain the 
modification of the angular distribution~\cite{Greco:PLB_90_2003,Hwa:PRC70_2004,
Solana:JPCS27_2005,Soalana:JPG34_2007,Stoecker:NPA750_2005,
Koch:PRL:96_2006,Majumdar:PRC73_2006,Muller:PLB618_2005,Ruppert:NPA774_2006},
the formation 
of Mach cones~\cite{Solana:JPCS27_2005,Soalana:JPG34_2007,Stoecker:NPA750_2005}, Cerenkov 
radiation~\cite{Koch:PRL:96_2006,Majumdar:PRC73_2006} etc., which can therefore 
be of phenomenological interest for the present experimental program at RHIC, LHC.

In addition, the jets, while traversing through the QGP,  will disturb the 
charge distribution around it and create the spikes in the induced 
charge density, hence in the screening potential, known as wakes, 
which can reveal the properties of medium. The wakes in both induced charge 
and current densities were first investigated by Ruppert et al.
\cite{Muller:PLB618_2005} within the framework of linear response theory
for a simple minded hot QCD medium for both weakly coupled and 
strongly coupled descriptions in the HTL calculations and in
the theory for quantum liquid, respectively.
In the same spirit, the wakes in the charge density and the 
potential was subsequently studied by Chakraborty et 
al.~\cite{Munshi:PRD74_2006}. The wake potential is discussed
in the literature by incorporating varieties of physical effects,
{\em viz.} the nonabelian effects from the 
resummation calculation \cite{Jiang:NPA832_2010}, 
the collisional effects~\cite{Munshi:JPG34_2007}, the momentum 
anisotropy~\cite{Roy:PRD86_2012,Roy:PRD88_2013}, 
the viscous effects~\cite{Jiang:NPA856_2011}, 
strongly-coupled nature of the medium in AdS/CFT 
framework~\cite{Liu:PRD93_2016} etc.

All the studies discussed hereinabove are suited for the fully central 
collisions, where the net baryon density is negligibly small
and the chance for producing strong magnetic field is very bleak, 
due to the symmetric configurations in central collisions.
In reality, a tiny fraction of events are truly central, in fact, most 
collisions occur with a nonzero impact parameter or centrality. 
When the two highly charged ions with extreme relativistic 
velocities collide noncentrally, an extremely large magnetic field ranging 
from $m_{\pi}^2$ ($\simeq 10^{18}$ Gauss) to 15 $m_{\pi}^2$ are
expected to be produced at RHIC and LHC, respectively, at the very 
early stages of the collisions~\cite{Skokov:IJMPA24_2009,Voronyuk:PRC83'2011}.
However, earlier works~\cite{Tuchin:AHEP2013'2013,Mclerran:NPA929_2014} 
did not emphasize much on the effects of {\em strong} magnetic field 
especially produced in relativistic heavy-ion collisions on
thermal QCD
because the life-span of the strong magnetic field was too short to have
any observable consequences in the heavy-ion phenomenology. They explained
the short span of the strong magnetic field by the fact that the time (at
the very early stages of the collisions) at which the magnetic field
could be produced, the medium is yet to form, as a  result the magnetic
field decays in vacuum very fast.
However some recent theoretical calculations
predict that a thermal medium could have produced as early as the
production of magnetic field so the electrical transport properties of the
medium plays a vital role about the longevity of the magnetic
field~\cite{Rath:PRD100'2019,Mclerran:NPA929_2014}. Consequently many 
theoretical works have started emerging to explore 
the effects of strong and long-lived magnetic field on 
QCD phenomena~\cite{Fukushima:PRD78'2008,
Braguta:PRD89'2014,Kharzeev:PRL106'2011,Gusynin:PRL73'1994}, which could
be verified experimentally in relativistic heavy ion
collisions at RHIC and LHC. Thus we get motivated to explore the effect 
of strong magnetic field on the abovementioned studies of 
wakes in charge density and potential because the jets are also 
produced at the time when magnetic field is produced.
For that purpose we 
have first analyzed the response of the medium due to the passage of
moving partons through the medium in the presence of strong magnetic field
by the dielectric function, whose evaluation is made complete by
the estimation of the gluon self-energy in the similar environment.
Once the tools are ready, we explore the effects of strong magnetic field 
on the wakes. The first noticeable observation is that 
the oscillation in the (scaled) induced charge density due to the faster
partons becomes less pronounced in the presence of strong magnetic field
whereas for smaller parton velocity, the effect is minimal. 
The second observation in the (scaled) wake potential 
for very fast moving partons along the parallel direction 
in the backward region is that the large depth of negative minimum of the 
Lennard-Jones (LJ) potential is reduced drastically, leaving the 
forward region almost 
unaffected, except that the screening now becomes stronger.
The third observation for the same in the 
perpendicular direction is that in both forward and
backward regions, the depth of LJ type potential gets
decreased severely, but still retains the forward-backward symmetry, 
as in the case for no magnetic field.
In brief, the wakes in the induced
charge density and in the potential as well have been changed significantly in
the presence of strong magnetic field. Therefore this observation may serve as a 
signal of the existence of strong magnetic field by the revisit of the correlations 
between high $p_T$ secondary particles in the pseudorapidity and azimuthal plane.

Our work proceeds as follows: First we have discussed the dielectric 
response due to the passage of the moving partons in a thermal QCD in 
an ambience of strong and homogeneous magnetic field in 
Section 2, wherein we have revisited the form of the gluon self-energy 
tensor at finite temperature ($T$) and finite magnetic field ($B$), and 
then calculate both real and imaginary parts of its longitudinal component 
at finite $T$ and strong $B$ in subsection 2.1. Hence the real and imaginary 
parts of complex dielectric function have been computed in subsection 2.2, 
which in turn facilitate to see the effects of strong magnetic field on 
the induced charged density and the resulting wake potential in
the coordinate space, in subsections 3.1 and 3.2, respectively.
Finally we conclude in Section 4.

\section{Dielectric response of thermal QCD in an ambience of strong and 
homogeneous magnetic field}
If the external disturbance is very small then the dielectric response of 
the medium can be envisaged by the dielectric tensor, $\epsilon^{ij}$, whose
longitudinal and transverse 
components can be extracted by the linear response theory~
\cite{Weldon1982,Kapusta1989,LeBellac:2000}
\begin{eqnarray}
\epsilon^{\rm Longitudinal} (k_0, \textbf{k}) = 1- \frac{\Pi^{\rm Longitudinal} (k_0,\textbf{k})}
{{\textbf{k}}^2}
\label{dielectric_longitudinal},\\
\epsilon^{\rm Transverse} (k_0, \textbf{k}) = 1- \frac{ \Pi^{\rm Transverse} (k_0,\textbf{k})}
{k_0^2}
\label{dielectric_transverse},
\end{eqnarray}
where the respective self-energies, $\Pi^{\rm Longitudinal}$ and 
$\Pi^{\rm Transverse}$ of the medium are obtained by the self energy tensor, 
$\Pi^{\mu
\nu}$. However, we are interested in the former equation because 
the longitudinal equation is related to the density fluctuations
in the medium, namely the space-charge field, which is our aim. 
Thus, we need to construct the gluon self-energy tensor in
finite $T$ and strong $B$, which will then give 
the longitudinal component.

\subsection{Gluon self-energy tensor in finite T and strong B}
We will revisit here how to construct the general form of the gluon 
self-energy tensor ($\Pi^{\mu \nu}$) for an isotropic QCD medium at 
finite $T$ and finite $B$. For that, we 
first consider the absence of magnetic field and turn on the temperature 
through a heat bath to the vacuum. Since the heat bath generically 
defines a local rest frame, $u^\mu$, so the Lorentz 
invariance is broken. Therefore, compared to vacuum, a larger tensor basis 
is needed, which is conveniently constructed by the available 
four-vectors, $k^\mu$, $u^\mu$ and the tensor, $g^{\mu\nu}$ by the two 
orthogonal tensors, compatible with the physical degree of freedom, 
{\em namely}~\cite{Braaten:PRL64_1990,Kobes:PRD_48_1992}
\begin{eqnarray}
P^{\mu\nu}_L(k)&=&-\frac{k^0}{\textbf{k}^2}(k^\mu u^\nu+u^\mu k^\nu)
+\frac{1}{\textbf{k}^2}\left[ \frac{{(k^0)}^2}{k^2}k^\mu k^\nu+k^2 u^\mu 
u^\nu\right]
\label{longitudinal_projection},\\
P^{\mu\nu}_T(k)&=&-g^{\mu\nu}+\frac{k^0}{\textbf{k}^2}(k^\mu u^\nu+u^\mu k^\nu)
-\frac{1}{\textbf{k}^2}(k^\mu k^\nu+k^2 u^\mu u^\nu)
\label{transverse_projection}
~,\end{eqnarray}
such a way that they satisfy the 4-dimensional transversality condition 
- $k_\mu P^{\mu \nu}_{T,L}=0$. 
The superscripts $T$ and $L$ represent the transverse
and longitudinal with respect to the three-momentum ($\mathbf{k}$), 
respectively. Thus the self-energy tensor at finite temperature is written
in the above basis
\begin{eqnarray}
\Pi^{\mu\nu}(k_0,\mathbf{k})=P^{\mu\nu}_T\Pi^T(k_0,\mathbf{k})+P^{\mu\nu}_L
\Pi^L(k_0,\mathbf{k}),
\label{self_energy_temperature}
\end{eqnarray}
where the structure factors, $\Pi^T$ and $\Pi^L$ are obtained by evaluating
the quark-loop, gluon-loops with 2-gluon and 
3-gluon vertices and ghost-loop contributions in HTL approximation
with the temperature as the hard scale for the loop
momenta~\footnote{In pure thermal medium,
the gluon loop-momenta may be hard, $\mathcal{O}$($T$) or soft, 
$\mathcal{O}$($g^{\prime}T$) because the Matsubara frequencies ($2m \pi T$) 
are integral multiples of $2\pi{T}$. Thus, the hard (soft)-momentum regime
requires $m \neq 0$ ($m=0$) bosonic matsubara mode. On the other hand the
quark loop-momenta will always be hard, even for $m=0$
Matsubara fermionic mode.} for both gluons and quarks. 

Now the presence of the magnetic
field breaks the rotational symmetry in the system because it
introduces an anisotropy in space. Thus, one can define a new four vector $b^\mu$ which is associated with the electromagnetic field tensor $F^{\mu\nu}$. In the rest frame of the heat bath, i.e., $u^\mu=(1, 0, 0, 0)$, 
$b^\mu$ can be defined uniquely as projection of ${\tilde{F}}_{\mu\nu}$ 
along $u^\nu$,
\begin{equation*}
b_\mu=\frac{1}{2B}\epsilon_{\mu\nu\rho\lambda}u^\nu F^{\rho\lambda}=\frac{1}{B}u^\nu{\tilde{F}}_{\mu\nu}=(0,0,0,1).
\end{equation*}
Hence, much larger basis is needed and can be 
constructed with the vectors, $k^\mu$, $u^\mu$, $b^\mu$ and the tensor,
$g^{\mu \nu}$. Therefore, in addition to $P^{\mu \nu}_T$ and 
$P^{\mu \nu}_L$ at finite temperature, two more tensors, 
$P^{\mu \nu}_\parallel$ and $P^{\mu \nu}_\perp$
are recently constructed in 
\cite{Hattori:PRD97_2018,Hattori:ANP:330_2013} 
for the leading-order perturbation theory
\begin{eqnarray}
P^{\mu\nu}_\parallel(k)&=&-\frac{k^0k^z}{k^2_\parallel}
(b^\mu u^\nu+u^\mu b^\nu)+\frac{1}{k^2_\parallel}
\left[(k^0)^2b^\mu b^\nu+(k^z)^2u^\mu u^\nu\right]
\label{prallel_projection},\nonumber\\
&=&-\left(g^{\mu\nu}_{\parallel}-\frac
{k_{\parallel}^\mu k_{\parallel}^\nu }
{k_\parallel^2}\right),\\
P^{\mu\nu}_\perp(k)&=&\frac{1}{k^2_\perp}\left[
-k^2_\perp g^{\mu\nu}+k^0(k^\mu u^\nu+u^\mu k^\nu)
-k^z(k^\mu b^\nu+b^\mu k^\nu)
+k^0k^z(b^\mu u^\nu+u^\mu b^\nu)\right.\nonumber\\
&& \left.-k^\mu k^\nu+(k^2_\perp-(k^0)^2)u^\mu u^\nu 
-\textbf{k}^2 b^\mu b^\nu\right]\label{perpendicular_projection},\nonumber\\
&=&-\left(g^{\mu\nu}_{\perp}-\frac
{k_{\perp}^\mu k_{\perp}^\nu }
{k_\perp^2}\right),
\end{eqnarray}
by demanding the same transversality condition for all tensors. The 
notations used in the above equations are as follows:
\begin{eqnarray*}
&&
g^{\mu\nu}_\parallel= {\rm diag} (1,0,0-1),~g^{\mu\nu}_\perp=
{\rm diag} (0,-1,-1,0),~
k^2_\parallel=(k_0)^2-(k_z)^2,~k^2_\perp=(k_x)^2+(k_y)^2.
\end{eqnarray*}

Unlike the thermal medium in the absence of magnetic field, where both quark and
gluon degrees of freedom have been treated in equal footing in hard thermal
loop approximation, quark and gluon degrees of freedom are treated
differently in presence of magnetic field because only quarks get affected
by the magnetic field and gluons remain unaffected. Thus in the presence
of strong magnetic field, the hard scale for the quark-loop momentum will 
be provided by the magnetic field whereas the hard scale for the 
gluon-loop momentum is still given by the temperature. As a result the quark
and gluon loops (denoted by $q$ and $g$, respectively) get separated in the 
general form of gluon self-energy tensor in finite $T$ and strong $B$ 
\begin{equation}
\Pi^{\mu\nu}(k)=P^{\mu\nu}_L\Pi^{g,L}(k)+P^{\mu\nu}_T\Pi^{g,T}(k)+
P^{\mu\nu}_\parallel\Pi^{q,\parallel}(k)+P^{\mu\nu}_\perp
\Pi^{q,\perp}(k)~,
\label{selfenergy_tensor}
\end{equation}
where the two new structure factors, $\Pi^{q,\parallel}$ and $\Pi^{q,\perp}$ 
give the contributions in the gluon self-energy due to the magnetic field 
only and their evaluation will be made from the quark loop and the thermal
contribution is still given by $\Pi^{g,L}$ and $\Pi^{g,T}$, whose
evaluations are done by the gluon (and ghost) loop only.

In the strong magnetic field, the quarks are constrained to be in the
lowest Landau level (LLL)($n=0$) only, which is witnessed by the vanishing
of the momentum transverse to the magnetic field ($p_\perp=0$)
in the dispersion relation ($E_0=\sqrt{p_z^2+m_f^2}$). This causes
the structure factor, $\Pi^{q,\perp}$ vanishingly small ($\Pi^{q,\perp} 
\approx 0$), therefore, the longitudinal component ($\mu=\nu=0$, denoted by $L^\prime$) of the gluon self-energy tensor 
at finite $T$ and strong 
magnetic field ($B$) in equation \eqref{selfenergy_tensor} is written in terms
of nonvanishing quark (q) and gluon (g) loops contributions
\begin{equation}
\Pi^{L^\prime} (k)=\Pi^{q,\parallel}(k)+\Pi^{g,L}(k),
\label{self_prime}
\end{equation}
because $P^{00}_T,P^{00}_\perp=0$.

We will now calculate the longitudinal component (denoted by $\parallel$) of 
the quark-loop contribution in the self-energy - $\Pi^{q,\parallel}$ in 
the real-time formalism. So we start with the ``11''-component 
of the quark-propagator matrix in thermal QCD in presence of strong magnetic 
field,
\begin{equation}
iS_{11} (p)=\Bigg[\frac{1}{{p_{\parallel}^2-m^2_f+
i\epsilon}}+2\pi in({p_0})\delta(p_{\parallel}^2-m^2_f)\Bigg]
(1+\gamma^{0}\gamma^{3}\gamma^{5})(\gamma^{0}p_{0}-\gamma^{3}
p_{z}+m_f) e^{\frac{-p_{\perp}^2}{\mid q_fB \mid}}.
\end{equation}
Hence the ``11"-component of the self-energy matrix due to
the quark-loop will be (omitting the label ``11")
{\small{
\begin{eqnarray}
{\Pi^{q,\mu\nu}} (k)&=& \frac{ig^2}{2} \sum_f \int\frac{d^{4}p}
{(2\pi)^4}{\rm Tr}[\gamma^{\mu}S_{11}(p)\gamma^{\nu}S_{11}(q)]\nonumber \\ 
&=&\frac{ig^2}{2} \sum_f \int\frac{d^{4}p}{(2\pi)^4}{\rm Tr}
[(\gamma^{\mu}(1+\gamma^{0}\gamma^{3}\gamma^{5})
(\gamma^{0}p_{0}-\gamma^{3}p_{z}+m_f)
\gamma^{\nu}(1+\gamma^{0}\gamma^{3}\gamma^{5})
(\gamma^{0}q_{0}-\gamma^{3}q_{z}+m_f)]\nonumber \\
&&\left\lbrace \frac{1}{p_{\parallel}^2-m_f^2+i\epsilon}
+2\pi i n(p_0)\delta(p_{\parallel}^2-m_f^2)\right\rbrace 
\left\lbrace \frac{1}{q_{\parallel}^2-m_f^2+i\epsilon}
+2\pi i n(q_0)\delta(q_{\parallel}^2-m_f^2)\right\rbrace
e^{\frac{{-p_{\perp}-q_{\perp}^2}}{|q_f|B}},
\label{pmn}
\end{eqnarray}
}}where the trace over the $\gamma$ matrices is calculated as
\begin{equation}
L^{\mu\nu}=8\left[p_{\parallel}^{\mu}q_{\parallel}
^{\nu}+p_{\parallel}^{\nu}q_{\parallel}^{\mu}
-g_{\parallel}^{\mu\nu}((p.q)_{\parallel}-m_f^2)
\right].
\label{trace}
\end{equation}

It is to be noted here that the coupling ($g$) 
in equation \eqref{pmn} depends on the magnetic field only because as mentioned above
the dominant scale for quarks in strong magnetic limit ($|q_fB| \gg T^2$)
is the scale related to magnetic field ($|q_fB|$). Moreover the 
coupling, $g$ depends only on the momentum longitudinal to the magnetic field
\footnote{because in strong magnetic field limit, only the lowest Landau level
(LLL) is occupied and hence the dispersion relation manifests 
the vanishing of the transverse component of momentum.} in the 
following manner~\cite{Ferrer:PRD91_2015} 
\begin{equation}
\alpha_{s}^\|(eB)=\frac{g^2}{4\pi}=\frac{1}{{\alpha_s^0(\mu_0)}^{-1}
+\frac{11N_c}{12\pi}\ln(\frac{k^2_z+M^2_B}{\mu_0^2})+\frac{1}{3\pi}
\sum_f \frac{|q_f B|}{\sigma}}, 
\end{equation}
where 
\begin{equation}
\alpha_s^0(\mu_0) = \frac{12\pi}
{11N_c\ln(\frac{(\mu_0^2+M^2_B)}{\Lambda_V^2})}.  
\end{equation}
where $k_z=0.1\sqrt{eB}$, $\mu_0=1.1$ GeV, $\Lambda_V=0.385$ GeV,
$M_B\sim~1$ GeV and the string tension, $\sigma=0.18 {\rm{GeV}}^2$. 

Due to the presence of magnetic field, the quark-loop momentum ($p$) 
becomes 
factorizable into parallel and perpendicular with respect to the
magnetic field. As a result, the tensor, $\Pi^{q,\mu\nu}(k)$ in
\eqref{pmn} also becomes factorizable,
{\em namely}
\begin{equation}
\Pi^{q,\mu\nu}(k)=\sum_f \Pi^{q,\mu\nu}(k_{\parallel})A(k_{\perp})~,
\label{selfenergy}
\end{equation}
where the perpendicular component of loop momentum ($p_\perp$) is 
integrated out to give rise the factor, $A (k_\perp)$
\begin{eqnarray}
A(k_{\perp})=\frac{\pi |q_{f}|B}{2}e^{-\frac{k_{\perp}^2}{2|q_{f}|B}},
\end{eqnarray}
which, in SMF approximation ($\exp (-\frac{k_{\perp}^2}{2|q_{f}|B} )
\approx 1$), becomes 
\begin{eqnarray}
A(k_{\perp})=\frac{\pi |q_{f}|B}{2}.
\label{pitr}
\end{eqnarray}

The part of the self energy that depends only on the parallel component 
of momentum, $\Pi^{q,\mu\nu}(k_{\parallel})$ due to 
quark-loop is decomposed into vacuum and 
medium components, {\em say}
\begin{equation}
\Pi^{q,\mu\nu}(k_\parallel) \equiv \Pi^{q,\mu\nu}_V(k_\parallel)+
\Pi^{q,\mu\nu}_n(k_\parallel)+\Pi^{q,\mu\nu}_{n^2}(k_\parallel),
\end{equation}
where the vacuum and thermal contributions due to single and double 
distribution functions, respectively, are 
\begin{eqnarray}
\Pi^{q,\mu\nu}_{V}(k_\parallel)&=&\frac{ig^2}{2(2\pi)^4}
\int dp_0 dp_z L^{\mu\nu}\left\lbrace
\frac{1}{(q_{\parallel}^2-m_f^2+i\epsilon)}\frac{1}
{(p_{\parallel}^2-m_f^2+i\epsilon)}\right\rbrace,
\label{pmnvac}\\
\Pi^{q,\mu\nu}_n(k_\parallel)&=&\frac{ig^2(2\pi i)}
{2(2\pi)^4}\int dp_0 dp_z L^{\mu\nu}\left\lbrace 
n(p_0)\frac{\delta(p_{\parallel}^2-m_f^2)}{(q_{\parallel}^2
-m_f^2+i\epsilon)}
+n(q_0)\frac{\delta(q_{\parallel}^2-m_f^2)}
{(p_{\parallel}^2-m_f^2+i\epsilon)}\right
\rbrace, 
\label{Pin1}\\
\Pi^{q,\mu\nu}_{n^2}(k_\parallel)&=&\frac{ig^2}{2(2\pi)^4}
\int dp_0 dp_z L^{\mu\nu}\lbrace 
(-4\pi^2)n(p_0)n(q_0)\delta(p_{\parallel}^2-m_f^2)
\delta(q_{\parallel}^2-m_f^2)\rbrace~. 
\label{Pin21}
\end{eqnarray}

Now we obtain the-real part of the vacuum term \eqref{pmnvac} as
\begin{eqnarray}
\Re {\Pi^{q,\mu\nu}_V(k_\parallel)}=\Big(g_{\parallel}^{\mu\nu}
-\frac{k_{\parallel}^{\mu}k_{\parallel}^{\nu}}{k_{\parallel}^2}
\Big)\frac{g^2}{2\pi^3} 
\left[\frac{2m_{f}^2}
{k_{\parallel}^2}
\left(1-\frac{4m_{f}^2}{k_{\parallel}^2}\right)^{-\frac{1}{2}}
\left\lbrace \ln\frac{{\Big(1-\frac{4m_{f}^2}
{k_{\parallel}^2}\Big)}^{1/2}-1}
{{\Big(1-\frac{4m_{f}^2}{k_{\parallel}^2}
\Big)}^{1/2}+1}\right\rbrace +1\right],
\end{eqnarray}
which was also calculated in \cite{Hattori:ANP:330_2013,Gusynin:NPB462_1996,
Fukushima:PRD83_2011, Hattori:ANP:334_2013,Mujeeb:EPJC77'2017}.  
Thus after multiplying the transverse momentum factor
\eqref{pitr}, the longitudinal component of the real-part of   
the vacuum term for massless quarks $(m_{f}=0)$ is obtained
in the simplified form 
\begin{equation}
\Re {\Pi^{q,\parallel}_V(k_0,k_z)}=\frac{g^2}{4\pi^2}
\sum_{f}\mid q_{f}B \mid 
\frac{k_{z}^2}{k_{\parallel}^2},
\label{Pi0vacuum}
\end{equation}
whereas the imaginary-part vanishes
\begin{equation}
\Im {\Pi^{q,\parallel}_V(k_0,k_z)}=0.
\end{equation}

Now we will calculate the medium contribution due to single 
distribution function, so we start with the real-part of the longitudinal
component from (\ref{Pin1}) as
\begin{eqnarray}
\Re {\Pi^{q,\parallel}_n (k_\parallel)}&=&-\frac{g^2}{2(2\pi)^3}
\int dp_0 dp_z L^{00}\Bigg[
n(p_0)\frac{\Big\{\delta(p_{0}-\omega_{p})
+\delta(p_{0}+\omega_{p})\Big\}}{2 \omega_p (q_{0}^2-q_{z}^2-m_f^2
)}\nonumber \\
&&+~n(q_0)\frac{\Big\{\delta(q_{0}-\omega_{q})
+\delta(q_{0}+\omega_{q})\Big\}}{2 \omega_q (p_{0}^2-p_{z}^2-m_f^2
)}\Bigg],
\label{pn00}
\end{eqnarray}
with the following notations:
\begin{eqnarray*}
L^{00}&=&8[p_{0}q_{0}+p_{z}q_{z}+m_f^{2}],\\
\omega_p&=&\sqrt{p_z^2+m_f^2},\\
\omega_q&=&\sqrt{q_z^2+m_f^2}=\sqrt{(p_z-k_z)^2+m_f^2}.
\end{eqnarray*}
After performing the $p_{0}$ integration first, the above Eq.(\ref{pn00})
becomes a one-dimensional equation in $p_z$
\begin{eqnarray}
\Re {\Pi^{q,\parallel}_n (k_\parallel)}&=&-\frac{g^2}{2(2\pi)^3}
\int dp_z\Bigg[\frac{L^{00}(p_0=\omega_p)~n(p_0=\omega_p)}{2\omega_p[(\omega_p-k_0)^2-\omega_q^2]}+\frac{L^{00}(p_0=-\omega_p)~n(p_0=-\omega_p)}
{2\omega_p[(\omega_p+k_0)^2-\omega_q^2]}\nonumber \\
&&+~\frac{L^{00}(p_0=\omega_q+k_0)~n(q_0=\omega_q)}{2\omega_q[(\omega_q+k_0)^2-\omega_p^2]}+\frac{L^{00}(p_0=-\omega_q+k_0)~n(q_0=-\omega_q)}{2\omega_q
[(\omega_q-k_0)^2-\omega_p^2]}
\Bigg] \label{pin}.
\end{eqnarray}
We will now evaluate the $p_z$ integration in the massless limit ($m_f=0$).
So, the following quantitites are calculated for $m_f=0$ as
\begin{eqnarray*}
L^{00}(p_0=p_z)&=&8[2p_z^2-p_zk_0-p_{z}k_{z}],\\
L^{00}(p_0=p_z)&=&8[2p_z^2+p_zk_0-p_{z}k_{z}],\\
L^{00}(p_0=q_z+k_0)&=&8[2p_z^2+k_z^2-3p_zk_z+q_zk_0],\\
L^{00}(p_0=-q_z+k_0)&=&8[2p_z^2+k_z^2-3p_zk_z-q_zk_0],
\end{eqnarray*} 
and then substituting them in Eq.\eqref{pin}, the medium term becomes 
in the massless limit
\begin{eqnarray}
\Re {\Pi^{q,\parallel}_n (k_\parallel)}&=&-\frac{4g^2}{2(2\pi)^3}
\int dp_z\Bigg[\frac{(2p_z-k_0-k_z)~n(p_0=p_z)}{[(p_z-k_0)^2-(p_z-k_z)^2]}+\frac{(2p_z+k_0-k_z)~n(p_0=-p_z)}{[(p_z+k_0)^2-(p_z-k_z)^2]}\nonumber\\&&+\frac{(2p_z^2+k_z^2-3p_zk_z+q_zk_0)~n(q_0=q_z)}{[(q_z+k_0)^2-p_z^2]~q_z}
+\frac{(2p_z^2+k_z^2-3p_zk_z-q_zk_0)~n(q_0=-q_z)}{[(q_z-k_0)^2-p_z^2]~q_z}
\Bigg].~~~~
\end{eqnarray}
After further simplification, the above equation yields into
a nicer form 
\begin{eqnarray}
\Re {\Pi^{q,\parallel}_n (k_\parallel)} =-\frac{2g^2}{(2\pi)^3}
\int dp_z\Bigg[-\frac{n(p_0=p_z)}{(k_0-k_z)}+\frac{n(p_0=-p_z)}
{(k_0+k_z)}+\frac{n(q_0=q_z)}
{(k_0-k_z)}-\frac{n(q_0=-q_z)}{(k_0+k_z)}\Bigg].
\end{eqnarray}
After substituting the distribution function, the medium contribution 
for the quark-loop is found to vanish identically
\begin{eqnarray}
\Re {\Pi^{q,\parallel}_n (k_\parallel)}
&=&-\frac{4g^2}{(2\pi)^3}\frac{k_z T}{k_0^2-k_z^2}
\left(\int_{-\infty}^{\infty} dt\frac{1}{e^{t}+1}-\int_{-\infty}^{\infty} 
dt^{\prime}\frac{1}{e^{t^{\prime}}+1}\right), \nonumber\\
&=&0~.
\label{selfn_final}
\end{eqnarray}
This result thus demonstrates the results of (1+1) 
dimensional massless QED (also known as Schwinger model), where 
the medium does not permeate to the vaccum. 
The same thing happens herein above because the (3+1)
dimensional QCD (in quark-loop) in strong magnetic field 
gets reduced into a (1+1) dimensional (longitudinal)
QCD, {\em mimicking} the Schwinger model.

On the other hand, the imaginary part due to single distribution function 
\eqref{Pin1} also vanishes
\begin{equation}
\Im {\Pi^{q,\parallel}_n(k_0,k_z)}=0.
\end{equation}
The medium contribution involving the product of 
two distribution functions (\ref{Pin21}) does not 
contribute to the 
real-part of the self-energy, i.e. 
	\begin{equation}
	\Re {\Pi^{q,\parallel}_{n^2} (k_0,k_z)}=0,
	\label{self_energy_nn}
	\end{equation}
whereas the computation of the imaginary-part is effective boiled down 
to the problem of one-loop massless fermion loop in two dimension as
an artifact of strong magnetic field. Hence the mapping of two-dimensional
massless fermion into bosons, known as CJT 
formalism of bosonization~\cite{Mandelstam:PRD11'1975} helps to
calculate the above imaginary-part. The bosonization technique 
was earlier well documented in \cite{Smilga:PLB278'1992,Fukushima:JPG39'2012}. Very recently this technique is applied
to QCD in a strong magnetic field~\cite{Fukushima:PRD93'2016}, where
the imaginary-part is obtained from 
the retarded current-current correlator as
\begin{equation}
\Im {\Pi^{q,\parallel}_{n^2}(k_0,k_z)}=-\frac{g^2\sum_f|q_f|B}{8\pi}k_0
[\delta(k_0-k_z)+\delta(k_0+k_z)],
\end{equation}
where the factor $|q_fB|/8\pi$ is the transverse density of states for 
the LLL states and arises due to the decoupling of the transverse dynamics 
from the longitudinal dynamics of LLL states.
	
We have thus finally obtained the real-part of the quark-loop contribution
by the vacuum \eqref{Pi0vacuum} contribution only, {\em i.e.}
\begin{eqnarray}
\Re {\Pi^{q,\parallel} (k_0,k_z)}&=&\Re {\Pi^{q,\parallel}_{V} (k_0,k_z)},\nonumber\\
&=&\frac{g^2}{4\pi^2}\sum_{f} 
|q_{f}|B \frac{k_z^2}{k_{\parallel}^2}~,
\label{real_quark_presence}
\end{eqnarray}
and the imaginary-part of the quark-loop contribution 
is given by the $n^2$ contribution 
\begin{eqnarray}
\Im{\Pi^{q,\parallel} (k_0,k_z)}&=&\Im {\Pi^{q,\parallel}_{n^2} (k_0,k_z)},\nonumber\\
&=&\frac{-g^2\sum_f|q_f|B}{8\pi}k_0
[\delta(k_0-k_z)+\delta(k_0+k_z)].
\label{img_quark_presence}
\end{eqnarray}
The above real and imaginary parts have also been addressed earlier in~\cite{Mustafa:PRD94'2016}. Both of these real- and imaginary-parts of quark-loop contribution are indeed devoid of medium (temperature) correction 
which out to be for the dynamics of massless quarks 
in (1+1) dimension.

Now let us calculate the remaining contribution by 
the gluon-loop, $\Pi^{g,L}$ in \eqref{self_prime}. Since 
the gluons are unaffected by the magnetic field so we 
calculate from the HTL pertubation theory, which is 
calculated earlier in~\cite{Braaten:NPB337_1990,
Pisarski:PRL63_1989,Thoma:arxiv0010164}, and is given by 
\begin{eqnarray}
\Pi^{g,L}(k_0,\textbf{k})&=&{g^\prime}^2 T^2\left(\frac{k_{0}}
{2|\textbf{k}|}\ln\frac{k_{0}+|\textbf{k}|+i\epsilon}{k_{0}
-|\textbf{k}|+i\epsilon}-1\right)~,
\label{pi_gluon}
\end{eqnarray}
whose real- and imaginary-parts can be extracted by the identity: 
\begin{equation}
\ln\left(\frac{k_{0}+|\textbf{k}|\pm i\epsilon}{k_{0} -|\textbf{k}|\pm i\epsilon}\right)
=\ln\left|\frac{k_{0}+|\textbf{k}|}{k_{0}-|\textbf{k}|}
\right|\mp i\pi\theta({\textbf{k}}^2-k^2_0),
\label{identity}
\end{equation}

Thus the real- and imaginary-parts are obatined from \eqref{pi_gluon} as
\begin{eqnarray}
\Re {\Pi^{g,L}(k_0,\textbf{k})}&=&{g^\prime}^2 T^2 
\left(\frac{k_{0}}{2|\textbf{k}|}\ln\left|\frac
{k_{0}+|\textbf{k}|}{k_{0}-|\textbf{k}|}\right|-1\right),
\label{real_gluon_absence}\\
\Im {\Pi^{g,L}(k_0,\textbf{k})}&=&-{g^\prime}^2 T^2\frac{\pi k_0}{2|\textbf{k}|},
\label{img_gluon_absence}
\end{eqnarray}
where the strong coupling, $g^\prime$ runs only with the temperature
({\em unlike} $g$), and its one-loop expression is
given by~\cite{Laine:JHEP'2005}   
\begin{eqnarray}
\alpha^{\prime}(T)=\frac{{g^\prime}^2(T)}{4\pi}=\frac{6\pi}{(33-2N_f)
\ln(\frac{2\pi T}{\Lambda_{QCD}})}~,
\end{eqnarray}
with $N_f$ is the number of flavour and $\Lambda_{QCD}\sim 0.200 GeV.$

Thus the real- and imaginary-part of the longitudinal component of 
the gluon self-energy tensor in finite $T$ in the presence of strong $B$
is obtained  by adding the finite $T$ correction
given by \eqref{real_gluon_absence} and \eqref{img_gluon_absence} 
and strong $B$ correction given by \eqref{real_quark_presence} and \eqref{img_quark_presence}
\begin{eqnarray}
\Re \Pi^{L^\prime}(k_0,\textbf{k}) &=& \Re \Pi^{g,L}(k_0,\textbf{k})+
\Re \Pi^{q,\parallel}(k_0,k_z), \nonumber\\
&=&{g^\prime}^2 T^2 
\left(\frac{k_{0}}{2|\textbf{k}|}\ln\left|\frac
{k_{0}+|\textbf{k}|}{k_{0}-|\textbf{k}|}\right|-1\right)
+\frac{g^2}{4\pi^2}\sum_{f} 
|q_{f}|B \frac{k_z^2}{k_{\parallel}^2},
\label{real_self_prime}\\
\Im \Pi^{L^\prime}(k_0,\textbf{k}) &=& \Im \Pi^{g,L}(k_0,\textbf{k})+
\Im \Pi^{q,\parallel}(k_0,k_z), \nonumber\\
&=&-{g^\prime}^2 T^2\frac{\pi k_0}{2|\textbf{k}|}-\frac{g^2\sum_f|q_f|B}
{8\pi}k_0[\delta(k_0-k_z)+\delta(k_0+k_z)]~,
\label{img_self_prime}
\end{eqnarray}
respectively.

\par To decipher the effects of strong magnetic field alone, we need to know
the abovementioned results of quark- and gluon-loops in the {\em absence} of 
magnetic field. The gluon-loop contribution is already mentioned above, and 
the quark-loop contribution can also be obtained from the gluon-loop 
expression in relativistic approximation, by replacing the color factor by the
flavour factor. For the sake of complete comparison, the real- and 
imaginary-part of quark-loop contribution are  
\begin{eqnarray}
\Re {\Pi^{q,L}(k_0,\textbf{k})}&=&{g^\prime}^2 T^2 
\left(\frac{N_f}{6}\right)\left(\frac{k_{0}}{2|\textbf{k}|}
\ln\left|\frac
{k_{0}+|\textbf{k}|}{k_{0}-|\textbf{k}|}\right|-1\right),
\label{real_quark_absence}\\
\Im {\Pi^{q,L}(k_0,\textbf{k})}&=&-{g^\prime}^2 T^2\left(\frac{N_f}{6}\right)
\frac{\pi k_0}{2|\textbf{k}|}
\label{img_quark_absence}.
\end{eqnarray}
Thus the real- and imaginary-part of the longitudinal
component of the gluon self-energy tensor in finite $T$
in the absence of $B$ is given by adding the gluon
loop contribution given by \eqref{real_gluon_absence} and
\eqref{img_gluon_absence} and quark loop
contribution given by \eqref{real_quark_absence} 
and \eqref{img_quark_absence} in absence of magnetic 
field 
\begin{eqnarray}
\Re {\Pi^{L}(k_0,\textbf{k})}&=&{g^\prime}^2 T^2 
\left(1+\frac{N_f}{6}\right)\left(\frac{k_{0}}{2|\textbf{k}|}
\ln\left|\frac
{k_{0}+|\textbf{k}|}{k_{0}-|\textbf{k}|}\right|-1\right),
\label{real_self_prime_absence}\\
\Im {\Pi^{L}(k_0,\textbf{k})}&=&-{g^\prime}^2 T^2\left(1+\frac{N_f}{6}\right)
\frac{\pi k_0}{2|\textbf{k}|}
\label{img_self_prime_absence}~,
\end{eqnarray}
respectively.

Therefore, the real- and imaginary-part of the longitudinal component of 
gluon self-energy in the presence and absence of magnetic field 
thus obtained, facilitate to study analytically the dielectric response 
of a hot QCD medium in respective scenarios.

\subsection{Dielectric response function in presence of strong
magnetic field}
The definition \eqref{dielectric_longitudinal} give the 
longitudinal component of the dielectric response function, 
which becomes complex. The real 
\eqref{real_self_prime} 
and imaginary parts \eqref{img_self_prime} of the longitudinal component of 
the gluon self-energy tensor give the real and imaginary parts of 
longitudinal dielectric function in presence of strong magnetic field 
\begin{eqnarray}
\Re\epsilon^{L^{\prime}}_{B\neq0}(k_0,\textbf{k})&=&1+\frac{{g^\prime}^2T^2}
{{\textbf{k}}^2}
\left(1-\frac{k_{0}}{2|\textbf{k}|}\ln\left|\frac
{k_{0}+|\textbf{k}|}{k_{0}-|\textbf{k}|}\right|\right)
-\frac{g^2}{4\pi^2}\sum_{f} 
|q_{f}|B\frac{1}{{\textbf{k}}^2}\frac{k_z^2}{k_{\parallel}^2}~,
\label{real_dielectric_mag}\\
\Im\epsilon^{L^{\prime}}_{B\neq0}(k_0,\textbf{k})&=&\frac{{g^\prime}^2T^2}
{{\textbf{k}}^2}\frac{\pi k_0}{2|\textbf{k}|}+\frac{g^2\sum_f|q_f|B}{8\pi}\frac{k_0}
{\textbf{k}^2}[\delta(k_0-k_z)+\delta(k_0+k_z].
\label{img_dielectric_mag}
\end{eqnarray}
Before moving on to the calculations of the induced charge density and the wake potential in the upcoming section, it is worth to mention here that due to the constraint - 
$k_0=\textbf{k} \cdot \textbf{v}$ in the Dirac delta function, that will appear in the charge density \eqref{external_charge} and for the choice $k_z=\textbf{k} \cos\theta$,  
the quark-loop contribution in the second term of equation 
\eqref{img_dielectric_mag}  
is only nonvanishing for $v=1$ ($v=|\textbf{v}|$). In our work, 
since we have 
considered the parton velocities less than one ($v=0.55, 0.99$)
so the imaginary part for quark-loop contribution will 
give the vanishing contribution.

To see the effects of strong magnetic field alone, the 
real and imaginary parts of the same in the absence of magnetic field
needs to be mentioned here as a baseline for comparison. These are
obtained from the HTL techniques~ \cite{Braaten:NPB337_1990,
Pisarski:PRL63_1989,Thoma:arxiv0010164}
\begin{eqnarray}
\Re\epsilon^{L}_{B=0}(k_0,\textbf{k})&=&1+\frac{{g^\prime}^2T^2}
{{\textbf{k}}^2}\left(1+\frac{N_f}{6}\right)
\left(1-\frac{k_{0}}{2|\textbf{k}|}\ln\left|\frac
{k_{0}+|\textbf{k}|}{k_{0}-|\textbf{k}|}\right|\right),
\label{real_dielectric_thermal}\\
\Im\epsilon^{L}_{B=0}(k_0,\textbf{k})&=&\frac{{g^\prime}^2T^2}
{{\textbf{k}}^2}
\left(1+\frac{N_f}{6}\right)\frac{\pi k_0}
{2|\textbf{k}|}.
\label{img_dielectric_thermal}
\end{eqnarray}

\vspace{3.2cm}
\begin{figure}[H]
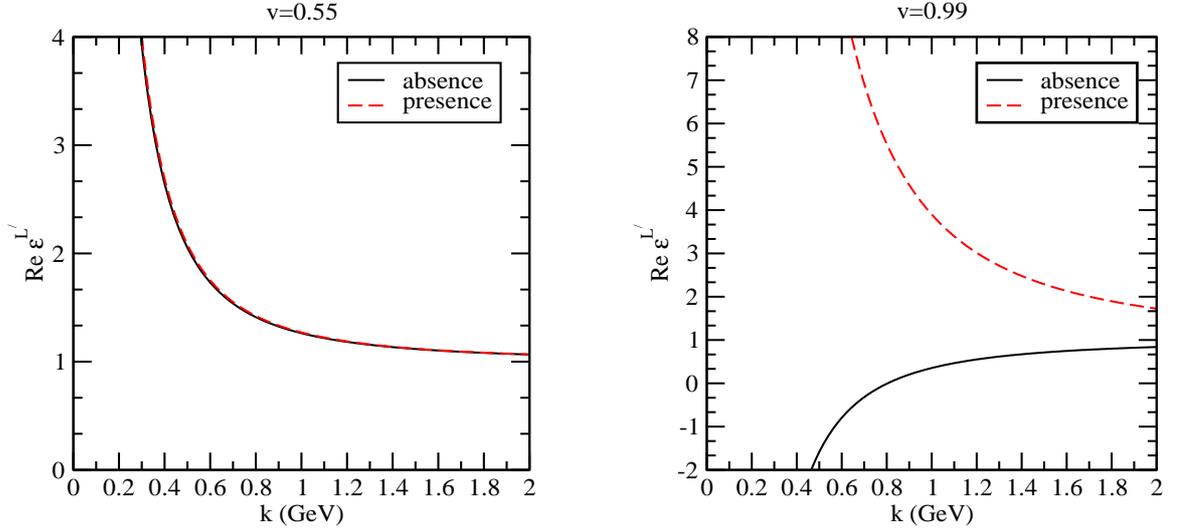

\begin{center}
\begin{tabular}{c c}
\includegraphics[width=7.5cm,height=7.5cm]{real55.eps}&~~~~~
\includegraphics[width=7.5cm,height=7.5cm]{real99.eps}\\
\end{tabular}
\caption{Real part of dielectric response function in absence
and presence of strong magnetic field for $v=0.55$ (left panel)
and $v=0.99$ (right panel), respectively.}
\label{real_dielectric}
\end{center}
\end{figure}
\vspace{2.5cm}
\begin{figure}[H]
\begin{center}
\begin{tabular}{c c}
\includegraphics[width=7.5cm,height=7.5cm]{img55.eps}&~~~~~
\includegraphics[width=7.5cm,height=7.5cm]{img99.eps}\\
\end{tabular}
\caption{Imaginary part of dielectric response function in absence
and presence of strong magnetic field for $v=0.55$ (left panel)
and $v=0.99$ (right panel), respectively.}
\label{img_dielectric}
\end{center}
\end{figure}
\nd In order to see the qualitative variations of real and imaginary parts of dielectric function obtained in Eqs.\eqref{real_dielectric_mag}-\eqref{img_dielectric_thermal}, with the parton velocity $v$, we have used the constraint put by the Dirac delta function defined in Eq.\eqref{external_charge} and $k_z=\textbf{k}\cos\theta$. In addition to this, we have taken the vectors $\textbf{v}$ and $\textbf{k}$ along the same direction means, $\cos\theta=\chi=1$, where the parton velocity $\textbf{v}$ is assumed to be in the z-direction.
Thus, to see the effects of magnetic field on the dielectric function, we have computed numerically its real and imaginary parts in the absence and presence of strong magnetic for slow ($v=0.55$) and 
fast ($v=0.99$) moving partons in Fig.\ref{real_dielectric} and 
Fig.\ref{img_dielectric}, respectively. 
For $v=0.55$, the change in 
the real part is meagre as compared 
to the case of no magnetic field. On the other hand, for $v=0.99$, its 
magnitude becomes much larger and the variation also becomes 
opposite compared to its counterpart in the absence of magnetic field.
The above anomalous behaviour of the real part of dielectric function 
can be understood physically in the following way: In the absence of 
magnetic field both quark and gluon loops in HTL approximation
have been treated on the same footing, apart from the color and flavour 
factors, so the momentum dependence for the real part of both loops 
(in equations \eqref{real_gluon_absence} and \eqref{real_quark_absence}) 
are the same which, in terms of velocity, 
look like ($\chi=\cos \theta$, $\theta$ is angle between $\textbf{v}$
and $\textbf{k}$, mentioned later in equation \eqref{external_charge})
\begin{eqnarray}
\Re {\Pi^{g (q),L}(v,\chi)}&\sim&
\left(\frac{v\chi}{2}\ln\left|\frac
{v\chi+1}{v\chi-1}\right|-1\right).
\end{eqnarray}
In presence of strong magnetic field, gluon loop is unaffected but the quark 
loop gets affected by the strong magnetic field (as seen in equation 
\eqref{real_quark_presence}), which in turn in terms of velocity looks as
\begin{eqnarray}
\Re {\Pi^{q,\parallel} (v)}
\sim \frac{1}{v^2-1},
\end{eqnarray}
which can be understood by the fact that the momentum integration 
in quark-loop becomes one dimensional as an artifact of phase-space 
reduction in strong
magnetic field, unlike the full three-dimensional 
momentum integration in the absence of strong magnetic field. As a 
result, the quark-loop contribution in the (real part) dielectric function
becomes much larger than its counterpart in the absence of
magnetic field for velocity ($v$) closer to $c$ ($v=0.99$).

More specifically, for small $|\textbf{k}|$, the real part 
becomes very large and approaches towards its counterpart at $B=0$, for 
large $|\textbf{k}|$. On the other hand, the imaginary part gets 
decreased for both slow and fast moving partons, due to the 
fact that the imaginary contribution due to quark-loop vanishes.
However, for larger parton velocity, its magnitude in both absence and presence
of magnetic field increases. In brief, only 
the real part of dielectric function for higher value of
parton velocity is largely affected, whose ramifications will now be explored
in both induced color charge density and wake potential in next section.

\section{Wakes in the presence of strong magnetic field}
When a moving test charge particle is traversing through the plasma, the 
charge density is induced locally, as a result a wake is observed in the
potential due to the induced charge density. Both of them depend on 
the velocity of the moving test charge as well as the distribution of 
background particles, which are going to be discussed in coming subsections.

\subsection{Induced color charge density}
Now we are going to discuss the wakes created in the induced charge density,
which can be expressed in terms of the dielectric response function 
$\epsilon^{L^\prime}(k_0,\textbf{k})$ and the external color charge density as 
\begin{eqnarray}
\rho_{\rm ind}^a(k_0,\textbf{k})=-\left\lbrace 1-\frac{1}{\epsilon^{L^\prime}
(k_0,\textbf{k})}\right
\rbrace\rho_{\rm ext}^a(k_0,\textbf{k}),
\label{c_induced}
\end{eqnarray}
where the external color charge density 
of charge $Q^a$ ($a=1,2, \cdot \cdot$) with velocity $\textbf{v}$ 
is given as 
\begin{eqnarray}
\rho_{\rm ext}^a(k_0,\textbf{k})=2\pi Q^a \delta(k_0-\textbf{k}\cdot\textbf{v}).
\label{external_charge}
\end{eqnarray}
Combining Eq.(\ref{c_induced}) and Eq.(\ref{external_charge}) and taking
the inverse Fourier transform, we obtain
the induced charge density in the coordinate space 
\begin{eqnarray}
\rho_{\rm ind}^a(\textbf{r},t)=2\pi Q^a\int\frac{d^3k}{{(2\pi)}^3}
\int\frac{dk_0}{2\pi}
e^{i(\textbf{k}\cdot\textbf{r}-k_0t)}\left\lbrace\frac{1}{\epsilon^{L^\prime}(k_0,\textbf{k})}-1\right
\rbrace \delta(k_0-\textbf{k}\cdot\textbf{v}).
\end{eqnarray}
Assuming that the partons are moving along the z-direction,
{\em i.e.} $\textbf{v} (0,0,v)$ and using the cylindrical 
coordinate system for $\textbf{r}(s, 0, z)$, spherical polar 
coordinate system for $\textbf{k} (k\cos\phi \sin\theta, k\sin\phi 
\sin\theta, \\k\cos\theta)$, we get the induced charge density
in terms of the spherical Bessel function of first kind ($J_0$)
\begin{eqnarray}
\rho_{\rm ind}^a(\textbf{r},t)=\frac{Q^a}{(2\pi)^2}\int_0^{\infty} k^2 
dk\int_{-1}^{1} d\chi J_{0}(ks\sqrt{1-\chi^2}) 
e^{i\Gamma}\left\lbrace\frac{1}{\epsilon^{L^\prime}(kv\chi,\textbf{k})}
-1\right\rbrace~,
\end{eqnarray} 
where the notations are defined by
\begin{eqnarray*}
\chi=\cos\theta,~\Gamma=k\chi(z-vt), \\
J_{0}(x)=\frac{1}{2\pi}\int_{0}^{2\pi} d\theta e^{ix\cos\theta}.
\end{eqnarray*}
By decomposing the dielectric function ($\epsilon^{L^\prime}$) into the real
and imaginary parts, the scaled induced charge density,
$\rho_{\rm ind}^0(\textbf{r},t)$ (=$
\frac{\rho_{ind}^a(\textbf{r},t)}{(Q^a/2\pi^2)m_D}$) 
is recast in the following form 
{\small{
\begin{eqnarray}
\rho_{\rm ind}^0(\textbf{r},t)=\int_0^{\infty} k^2 
dk\int_{0}^{1} d\chi J_{0}(ks\sqrt{1-\chi^2}) 
\left[\cos\Gamma\left\lbrace\frac{\Re\epsilon^{L^\prime}(kv\chi,\textbf{k})}
{\Delta}-1\right\rbrace
+\sin\Gamma\frac{\Im\epsilon^{L^\prime}(kv\chi,\textbf{k})}{\Delta}\right],
\label{chargedensity}
\end{eqnarray}}} 
with $\Delta={[\Re\epsilon^{L^\prime}(kv\chi,\textbf{k})]}^2+
{[\Im\epsilon^{L^\prime}(kv\chi,\textbf{k})]}^2$. 

\begin{figure}[H]
\begin{center}
\begin{tabular}{c c}
\includegraphics[width=9cm,height=10cm]{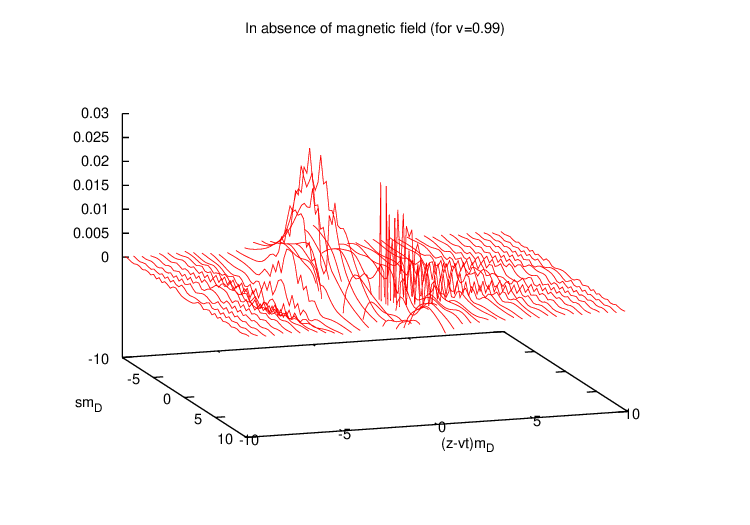}&~
\includegraphics[width=9cm,height=10cm]{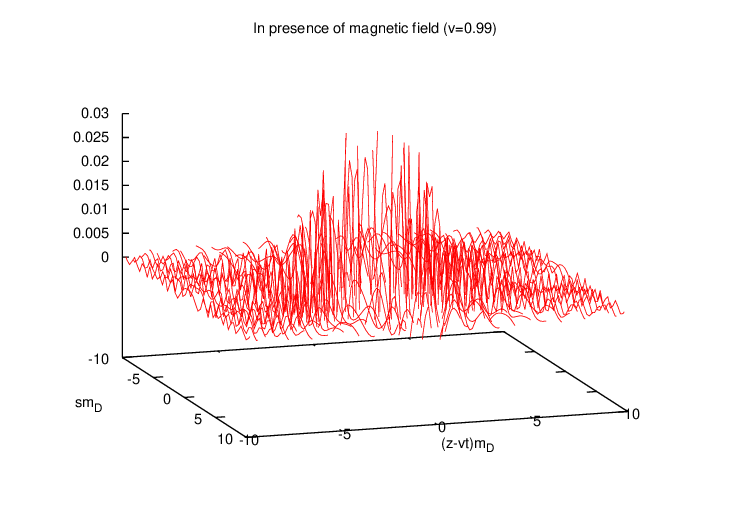}\\
\end{tabular}
\caption{Scaled induced charge density ($\rho_{\rm ind}^0$) in absence (Left panel) 
and presence (Right panel) of strong magnetic field for higher parton velocity $v=0.99$, 
respectively.}
\label{charge99}
\end{center}
\end{figure}

\begin{figure}[H]
\begin{center}
\begin{tabular}{c c}
\includegraphics[width=9cm,height=10cm]{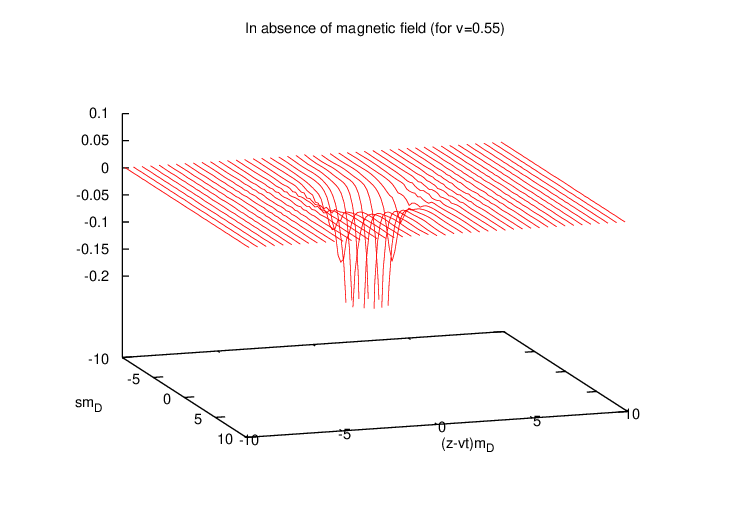}&~
\includegraphics[width=9cm,height=10cm]{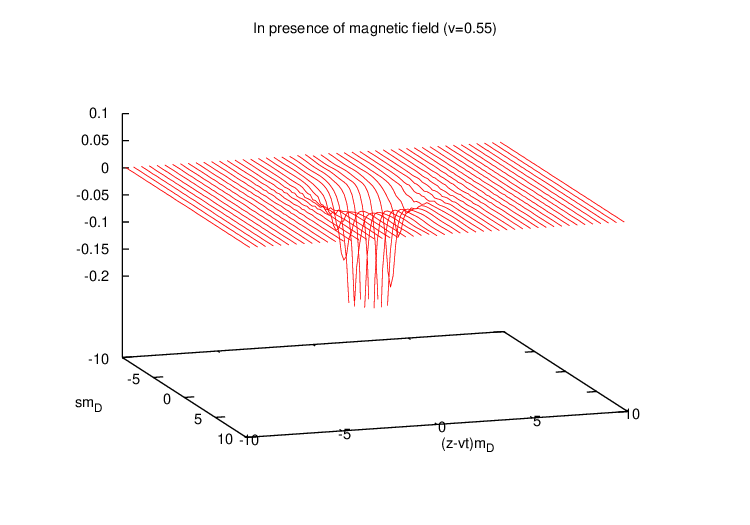}\\
\end{tabular}
\caption{Scaled induced charge density ($\rho_{\rm ind}^0$) in absence 
(Left panel) and presence (Right panel) of strong magnetic field for smaller
parton velocity $v=0.55$, respectively.}
\label{charge55}
\end{center}
\end{figure}

We are now going to investigate the effects of strong magnetic field on the
induced charge density quantitatively through the three-dimensional 
plots in Fig.\ref{charge99} and 
Fig.\ref{charge55} for 
higher and smaller parton velocities, respectively.

For faster (than the average phase speed, $v_p$) moving partons, 
the induced 
charge density is found to 
carry the oscillations along the direction of moving charge
(as seen in Fig.\ref{charge99}), which
generate the Cerenkov like radiation and Mach shock waves. This oscillatory 
behaviour is always found, but the presence of magnetic field dampens the 
oscillation much. Because the real part of dielectric response function 
gets enhanced due to the presence of strong magnetic field
(as illustrated in the right panel of Fig.\ref{real_dielectric}) and
results a decrease in the amplitude.

On the other hand, for smaller parton velocities ($v=0.55$), magnetic field 
does not alter the induced charge density significantly, because 
the real part of the dielectric function in the  similar case (seen in 
the left panel of Fig.\ref{real_dielectric}) does not change much in the 
presence of magnetic field. 

\subsection{Wake potential}
In this subsection we will study the wakes in the potential due to the 
induced charge density as discussed in the previous subsection.
The wake potential in the momentum space is obtained from the Poisson 
equation as 
\begin{eqnarray}
\Phi^a(k_0,\textbf{k})=\frac{\rho_{\rm ext}^a(k_0,\textbf{k})}
{k^2\epsilon^{L^\prime}(k_0,\textbf{k})}.
\end{eqnarray}
Substituting the external color charge density from \eqref{external_charge}, 
the inverse Fourier transform gives 
the wake potential in the coordinate space 
\begin{eqnarray}
\Phi^a(\textbf{r},t)=Q^a\int \frac{d^3k}{(2\pi)^3}\int e^{i(\textbf{k}\cdot\textbf{r}-\textbf{k}\cdot\textbf{v}t)}
\frac{1}{k^2 \epsilon^{L^\prime}(\textbf{k}\cdot\textbf{v},~\textbf{k})}.
\end{eqnarray} 
Similar to the coordinate transformations used for obtaining 
the induced charge density, we obtain the wake potential 
in the coordinate space
\begin{eqnarray}
\Phi^a(\textbf{r},t)=\frac{Q^a}{2\pi^2}\int_0^{\infty} 
dk\int_{0}^{1} d\chi J_{0}(ks\sqrt{1-\chi^2}) 
\left[\frac{\cos\Gamma\Re\epsilon^{L^\prime}(kv\chi,\textbf{k})}{\Delta}
+\frac{\sin\Gamma\Im\epsilon^{L^\prime}(kv\chi,\textbf{k})}{\Delta}\right],
\label{wakepotential}
\end{eqnarray} 
which will be computed for two directions, one is along the direction 
of motion of the partons $(\textbf{r} \parallel \textbf{v})$ and
other is perpendicular to the direction of motion of the partons 
$(\textbf{r} \perp \textbf{v})$. Therefore, the scaled wake potential along the 
direction of motion (we presumed that the partons are 
moving along the z-direction, i.e. $s=0$) is thus obtained as
\begin{equation}
\Phi^0_{\parallel}(\textbf{r},t)=\int_0^{\infty} dk\int_{0}^{1} d\chi  
\left[\frac{\cos\Gamma\Re\epsilon^{L^\prime}(kv\chi,\textbf{k})}{\Delta}
+\frac{\sin\Gamma\Im\epsilon^{L^\prime}(kv\chi,\textbf{k})}{\Delta}\right],
\label{wakepotential_parallel}
\end{equation} 
and the potential perpendicular to the direction of motion 
is ($\Gamma^{\prime}=k\chi vt$)
\begin{equation}
\Phi^0_{\perp}(\textbf{r},t)=\int_0^{\infty} dk\int_{0}^{1} d\chi  
J_{0}(ks\sqrt{1-\chi^2})
\left[\frac{\cos\Gamma^{\prime}\Re\epsilon^{L^\prime}(kv\chi,\textbf{k})}{\Delta}
-\frac{\sin\Gamma^{\prime}\Im\epsilon^{L^\prime}(kv\chi,\textbf{k})}{\Delta}\right].
\label{wakepotential_perpendicular}
\end{equation} 

\vspace{3cm}
\begin{figure}[H]
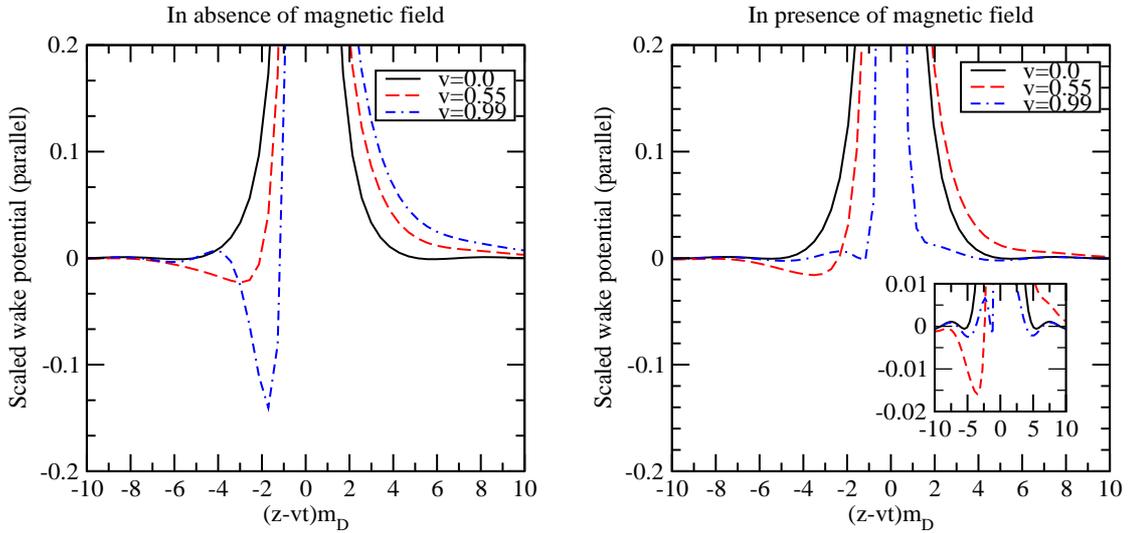

\begin{center}
\begin{tabular}{c c}
\includegraphics[width=8.0cm,height=8.0cm]{wake_tparall.eps}&~
\includegraphics[width=8.0cm,height=8.0cm]{wake_bparall.eps}\\
\end{tabular}
\caption{Left panel: Scaled wake potential ($\Phi^0_{\parallel}$) along the motion of 
the partons for smaller and higher parton velocities in the absence of magnetic field. 
Right panel: Same as the left panel but in the presence of magnetic field.}
\label{wake_parallel}
\end{center}
\end{figure}

\begin{figure}[H]
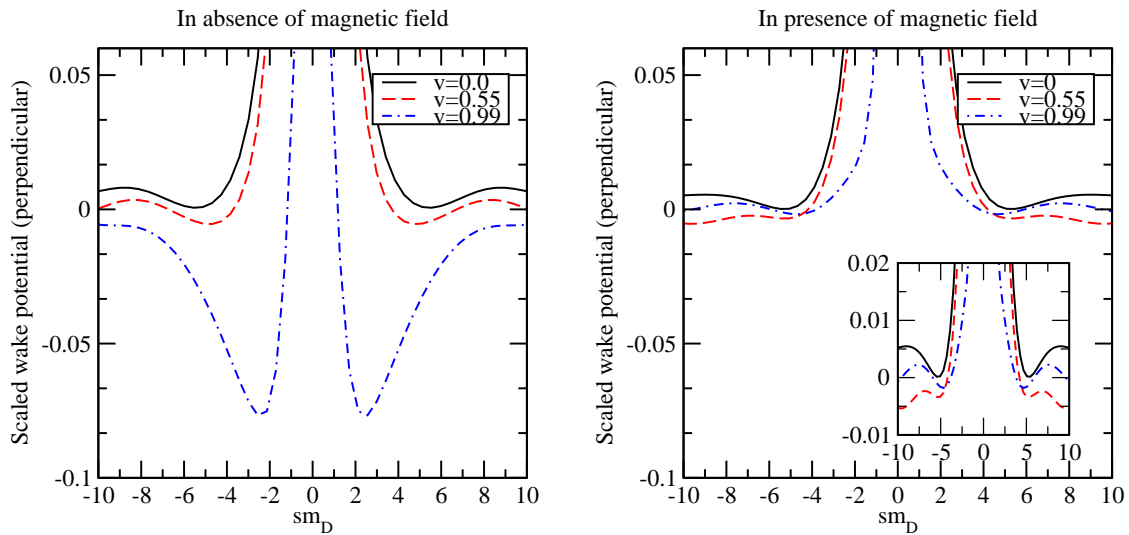

\begin{center}
\begin{tabular}{c c}
\includegraphics[width=8cm,height=8cm]{wake_tperp.eps}&~
\includegraphics[width=8cm,height=8cm]{wake_bperp.eps}\\
\end{tabular}
\caption{Left panel: Scaled wake potential ($\Phi^0_{\perp}$) perpendicular to the 
direction of motion of the partons for smaller and higher parton velocities 
in the absence of magnetic field. Right panel: Same as the left panel but 
in the presence of magnetic field.}
\label{wake_perpendicular}
\end{center}
\end{figure}

Finally, to visualize the effects of magnetic field, we have computed
the wake potential both in absence and presence of strong magnetic field 
along the parallel and perpendicular directions in Fig.\ref{wake_parallel} 
and Fig.\ref{wake_perpendicular}, 
respectively. In the absence of magnetic field (left panel of Fig.\ref{wake_parallel}), 
for the static case, the wake potential is 
by default forward backward symmetric but for finite $v$ it falls off very 
fast compared to the static one and loses the forward backward symmetry. 
In the backward region, the wake potential 
decreases with the distance and shows a minimum when 
partons move with the velocity $v=0.55$ (less than $v_p$).
However for parton velocity greater than $v_p$, i.e. $v=0.99$, 
the wake potential shows an oscillatory behavior and behaves 
differently from $v=0.55$ case and the width of the negative minimum
is increased and shifted towards the origin compared to $v=0.55$. 
Thus in the backward region, the wake potential is a
Lennard-Jones potential type which shows a short-range repulsive part
and a long-range attractive part. 
On the other hand, in the forward region, the wake potential behaves more like
a screened Coulomb potential and attains the Coulomb form on 
increasing the value of parton velocity. Thus the forward part is not
much affected by the motion of charged particles.
Now, in the presence of magnetic field (right panel of Fig.\ref{wake_parallel}), 
in the backward region the wake potential decreases 
on increasing the value of $(z-vt)m_D$, and shows a minimum 
but the depth of the minimum is in general decreased compared to 
the absence of magnetic field. For $v=0.55$ the width 
has been reduced slightly, however for $v=0.99$, the reduction
is very large. Moreover, in the backward region for $v=0.99$, the potential 
also shows oscillatory behavior but with lesser amplitude.
This can be understood again in terms of enhancement of real part of 
dielectric response function in presence of magnetic 
field (seen in Fig.\ref{real_dielectric}).
Thus in the backward region, the wake potential is still found to be of 
Lennard-Jones type potential. On the other hand in the forward region, 
the wake potential is screened Coulomb potential. However, for higher parton 
velocity, it does not attain the Coulomb form as found in the case when there 
is no magnetic field, rather it becomes more screened on increasing 
the value of parton velocity. Overall the magnetic field is found to affect 
the higher velocity partons only.

\vspace{3.3cm}

In absence of magnetic field, the wake potential, perpendicular to the moving 
partons (left panel of Fig.\ref{wake_perpendicular}), is 
symmetric in backward and forward regions, independent of the speed of the 
moving partons and the depth of negative minimum increases with
the velocity of parton.  The nature of wake potential is 
Lennard-Jones type for negative as well as positive value of 
$s m_D$. On the other hand, in the presence of strong magnetic
field (right panel of Fig.\ref{wake_perpendicular}), the potential has the 
forward backward symmetry and of Lennard-Jones nature. However, the magnetic field 
decreases the depth of negative minimum for higher parton velocity, 
which is opposite to the absence of magnetic field.

\section{Conclusions}
In the present study we have calculated the wakes 
in the induced charge density as well as in the potential generated
due to the passage of highly energetic partons through a thermal QCD medium 
in the presence of strong magnetic field, believed to be created in the off 
central events of heavy-ion collision experiments.  
For that purpose we have calculated the responses of the 
medium both in presence and absence of strong magnetic field through 
the dielectric function. To calculate the response function, first 
we have revisited the general covariant form for the one-loop gluon self-energy 
tensor at finite temperature and finite magnetic field and then approximated the 
relevant structure functions at finite temperature in the strong magnetic field limit. 
Thus we have obtained the real and imaginary parts of complex dielectric response function 
both in absence and presence of strong magnetic field for slow 
($v=0.55$) and fast ($v=0.99$) moving partons. For slow moving partons, 
we have found that the real part of dielectric response function 
is not much affected by the magnetic field whereas for fast moving
partons, it becomes very large compared to its counterpart in absence of magnetic field 
for small $|\textbf{k}|$, however, it approaches towards the value in 
absence of magnetic field for large $|\textbf{k}|$.  On the other hand the magnitude of the 
imaginary part of dielectric response function is slightly decreased for both
slow and fast moving partons in presence of strong magnetic field. 
Finally, we have computed the scaled induced charge density ($\rho^0_{\rm ind}$) 
and the scaled wake potential for parallel ($\Phi^0_\parallel$) and 
perpendicular ($\Phi^0_\perp$) directions to the motion of moving partons. 
The oscillation found in $\rho^0_{\rm ind}$ due to the very 
fast partons
becomes less pronounced in the presence of strong magnetic field
whereas for smaller parton velocity, no significant change is 
observed. The wake potential, $\Phi^0_\parallel$ for very fast 
moving partons is found to be of Lennard-Jones (LJ) type and the depth of 
negative minimum in the backward region is reduced drastically, resulting 
in the decrease of amplitude of oscillation due to the strong magnetic field.
On the other hand in the forward region, $\Phi^0_\parallel$ remains
the screened Coulomb one, except the screening now becomes
relatively stronger for higher parton velocity. For $\Phi^0_\perp$ in 
both forward and backward region, the depth of negative minimum in LJ 
potential gets decreased severely for fast moving partons, but still retains 
the forward-backward symmetry. However, for lower parton velocity,
the magnetic field does not affect the nature of $\Phi^0_\perp$ significantly. 


\section*{Acknowledgements}
BKP is thankful to the CSIR (Grant No.03 (1407)/17/EMR-II), Government of India for the
financial assistance.

\end{document}